\documentstyle[12pt]{article}
\begin{document}

\begin{center}
{\Large \bf KAON MASS IN DENSE MATTER}

\vspace{0.5in}

{\bf Abhijit Bhattacharyya$^a${\footnote{email: phys@boseinst.ernet.in}},
 Sanjay K. Ghosh$^a${\footnote{email: phys@boseinst.ernet.in}},
 S. C. Phatak$^b${\footnote{email: phatak@iopb.ernet.in}} and 
Sibaji Raha$^a${\footnote{email: sibaji@boseinst.ernet.in}}}

\vspace{0.5in}
a) Department of Physics, Bose Institute, \\ 93/1, A. P. C. Road, Calcutta
700 009, INDIA \\
b) Institute of Physics, Bhubaneswar 751 005, INDIA
\end{center} 
\begin{abstract}

The variation of kaon mass in dense, charge-neutral baryonic matter at 
beta-equilibrium has been investigated. The baryon interaction has been 
included by means of nonlinear Walecka model, with and without hyperons and 
the interaction of kaons with the baryons has been incorporated through the 
Nelson-Kaplan model. A self-consistant, one-loop level calculation has been 
carried out. We find that at the mean field level, the presence of the 
hyperons makes the density-dependence of the kaon mass softer. Thus, the kaon 
condensation threshold is pushed up in the baryon density. The loop diagrams 
tend to lower the kaon condensation point for lower values of $a_3 m_s$. We 
also find that the S-wave kaon-nucleon interaction plays the dominant role in 
determining the on-set of kaon condensation and the contribution of the P-wave 
interaction is insignificant. 
\end{abstract}



\vspace{1.5in}

The study of hadronic matter at large temperature and densities is a field of 
immense current interest because of the possibility of novel phases, such as 
pion condensation, kaon condensation, quark-matter formation etc, appearing at 
higher baryonic densities and temperatures, which may be achieved in 
relativistic heavy ion collisions and /or dense stars. One of the possible 
interesting features of the dense hadronic matter, investigated by several 
authors \cite{a,b,c,d,e,f,g}  of late, is that a charge-neutral, dense 
hadronic matter may undergo kaon condensation at sufficiently high densities 
( baryon densities $\sim $ 3-5 times  nuclear matter density ). If this is 
true, then one would expect that the cores of neutron stars may consist of 
kaon-condensed baryonic matter. The nuclear equation of state in the presence 
of kaon condensation being softer, this would affect the limits on neutron star
masses and radii. A good understanding of the density at which the kaon 
condensation may occur is also of interest in the context of heavy ion 
collisions aimed at looking for quark matter as this density is quite close to 
the density of the expected hadron to quark matter phase transition. Thus, one 
would like to know whether one passes through a kaon-condensed phase between 
the baryonic and the quark phases as the nuclear density is increased. The 
kaon-condensation threshold, {\it i.e.}, the density at which kaon-condensation 
begins, is given by the density at which the effective kaon mass in the 
nuclear medium equals the kaon chemical potential. For a system having zero 
strangeness, this threshold is equivalent to having zero kaon mass. Thus the 
investigation of the behaviour of the kaon mass as a function of baryon 
density is of importance.  

The possibility of kaon condensation in the nuclear matter was first 
investigated by Nelson and Kaplan\cite{a}. In their calculation these authors 
used a $SU(3) \times SU(3)$ Lagrangian ( see later for the details ). In their 
model the kaon condensation is predominantly driven by an attractive S-wave 
kaon-nucleon interaction. This should be contrasted with the pion condensation 
which is 
driven by the P-wave pion-nucleon interaction. The calculation of Nelson and 
Kaplan was done in the mean field approximation and the hyperons, which are 
likely to appear in the dense hadronic matter, were not included. The 
calculation of Nelson and Kaplan has been extended by a number of authors 
\cite{c,d,e,f,g}. In some of the extensions the nonlinear Walecka model 
\cite{g} has been used to incorporate the baryonic interactions and to employ 
a somewhat more realistic nuclear equation of state. The hyperons which are 
expected to appear at higher nuclear density are also included. However, most 
of these calculations are mean field calculations and do not include higher 
loop corrections. 

In the present work we have studied the properties of kaons in the nuclear
medium. In particular, we have computed the kaon propagator in the nuclear 
medium and determined the effective kaon mass in the nuclear medium from the 
pole position of  the kaon propagator. We go beyond the mean field 
approximation ( essentially Fig. 1(a) ) and include the oyster diagrams 
( Fig. 1(b) ) in our calculations. We also investigate the effect of P-wave 
kaon-baryon interaction on the kaon propagator ( fig. 1(c) ). 

Before going on to describing our calculation, let us first consider the two 
relevant Lagrangians referred to in the above. The Nelson-Kaplan $SU(3) \times
SU(3)$ Lagrangian is \cite{a,f}
\begin{eqnarray}
{\cal L }_{NK}&=&\frac{1}{4} {\rm Tr \;} \partial_{\mu}U\partial^{\mu}
U^{\dagger}
+c {\rm Tr \; } (m_{q}(U+U^{\dagger}-2)) 
+ {\rm Tr \;} \bar B \gamma^{\mu}i\partial_{\mu}B 
 +  i \; {\rm Tr \; } \bar B \gamma_{\mu}[V_{\mu}, B] \nonumber \\
 & + &  D \; {\rm Tr \;} \bar B\gamma_{\mu}\gamma_{5}\{A_{\mu}, B\}
 + F \; {\rm Tr \;} \bar B \gamma_{\mu} \gamma_{5}[A_{\mu}, B] 
+ a_{1} \; {\rm Tr \; } \bar B (\zeta m_{q} \zeta + h.c.)B \nonumber \\
 &  + & a_{2} \; {\rm Tr \; } \bar B B(\zeta m_{q} \zeta + h.c.)  
 +a_{3} \; {\rm Tr \;} (m_{q}U + h.c.) \; {\rm Tr \;} \bar B B ,
\label{eq0}
\end{eqnarray}
where $B$ is the baryon octet, $M$ is the pseudoscalar meson octet, $m_{q}$ is 
quark mass matrix, $f$ is the pion decay constant, $U =\exp (\sqrt{2}iM/f)$ 
and $\zeta^{2}=U$. The mesonic vector and axial vector currents ($ V^\mu $ and 
$ A^\mu $ respectively ) are given by, 
\begin{eqnarray}
V_{\mu}= \frac {1}{2}(\zeta^{\dagger}\partial_{\mu} \zeta + \zeta \partial
_{\mu} \zeta^{\dagger})  \\
A_{\mu}= \frac {1}{2}(\zeta^{\dagger}\partial_{\mu} \zeta - \zeta \partial
_{\mu} \zeta^{\dagger})
\label{eq1}
\end{eqnarray}

In our calculation, the quark masses are chosen to be $m_u = m_d = 0$ and 
$m_s = 150 MeV$ and the pion decay constant is chosen to be 93 MeV. Apart from 
these, the other parameters of the Lagrangian are the weak interaction 
constants $D = 0.81$, $F = 0.44$ ( with $g_A = F + D$ ) and $a_1$, $a_2$ and 
$a_3$. The parameters $a_1$ and $a_2$ are determined by fitting the nucleon, 
lambda and cascade masses. In terms of the strange quark mass ($m_s$), 
$a_{1}m_{s}=-64 MeV$ and $a_{2}m_{s}=-134 MeV$. The parameter $a_3$ is 
related to the strangeness content of the nucleon and kaon-nucleon sigma term 
$\Sigma^{kn}$ \cite{f}. It must be noted that $\Sigma^{kn}$ is not 
well-determined experimentally and therefore $a_3$ is not known particularly 
well. In our calculation, $a_3m_s$ has been varied from -134 MeV to  -310 
MeV. These correspond to zero and 20\% strangeness content of the nucleon, 
respectively.  

The nonlinear Walecka model Lagrangian is \cite{i}
\begin{eqnarray}
{\cal L}_W~&=&\sum_{i} \bar B _{i} (i\gamma^{\mu}\partial_{\mu}
- m_i +g_{\sigma i}
\sigma+ g_{\omega i} \omega_{\mu} \gamma^{\mu}- 
g_{\rho i} \rho^{a}_{\mu} \gamma^{\mu} T_{a} ) B_{i} 
-{1 \over {4}} \omega ^{\mu \nu} \omega_{\mu \nu} \nonumber \\ 
&+&{1 \over {2}} m^{2}_{\omega} \omega_{\mu} \omega^{\mu}  
+ {1 \over {2}} ( \partial_{\mu} \sigma \partial^{\mu} \sigma-
m^{2}_{\sigma} \sigma^{2}) 
- {1 \over {4}} \rho^{a}_{\mu \nu} \rho^{\mu \nu}_{a} +
{1 \over {2}} m_{\rho}^{2} \rho^{a}_{\mu} \rho^{\mu}_{a} \nonumber \\
 &-& {1 \over {3}} bm_{N} (g_{\sigma N} \sigma)^{3} -
{1 \over {4}} c( g_{\sigma N} \sigma )^4 +
\sum_{l}\bar\psi_{l}(i\gamma^{\mu}\partial_{\mu}-m_l)\psi_{l}.
\label{eq2}
\end{eqnarray}

Here $B, \; \sigma, \; \omega_\mu, \; \rho_\mu^a ,$ and $\psi_l$ are the 
baryon, sigma, omega, rho and lepton fields respectively, g's are the baryon
-meson coupling constants and b and c are the coefficients of the cubic and 
quartic $\sigma$ meson self interaction terms. The strangeness 0 and -1 
baryons and electrons and muons are included in the present calculation. The 
masses of rho and omega mesons are chosen to be their experimental values and 
the sigma mass and meson-baryon coupling constants are chosen by fitting the 
nuclear saturation properties\cite{h}. The coupling constants of the mesons 
with the strange baryons are not fixed by the nuclear saturation properties. 
We therefore use the quark model estimates for these\cite{i}.

The procedure adopted in the present calculataion is as follows. The equation 
of state of the hadronic matter is calculated from the Walecka model (4) in 
the mean field approximation. The calculation yields the densities of 
different baryon species and their effective masses as a function of baryon 
density. This information is then used in the calculation of the kaon 
propagator $D(\vec k, \omega)$ 
\begin{eqnarray}
D(\vec k, \omega) = \frac{1}{\omega^2 - k^2 - m_k^2} ( 1 + \Pi (\vec k, \omega)
D(\vec k, \omega) )
\end{eqnarray}
in the hadronic medium. The kaon self energy $\Pi(\vec k, \omega)$ is 
calculated, from the Lagrangian (1), by including the diagrams shown in Fig. 1. 
Thus, our calculation includes the diagrams upto one-loop level and upto fourth
power in $1/f$. The loop diagrams, which are divergent, have been properly 
regularised so as to yield the physical mass of Kaon at zero baryon density 
using standard dimensional regularisation techniques \cite{ia}. In Fig. 1, (a) 
and (b) are due to the S-wave interaction, whereas (c) is due to the P-wave 
interaction. For the sake of brevity we have not given the full expression of 
self energy here. This will be reported in a full paper elsewhere.   

The pole of the kaon propagator, as a function $\vec k$ gives the relation
between the energy and momentum of kaons in the medium. In particular, the 
value of the real part of $\omega$ which satisfies the equation \cite{j}
\begin{eqnarray}
D^{-1}(\vec k \rightarrow \vec 0, \omega) &=& \omega^2 - m_k^2 - Re[\Pi (
|{\vec k}|\rightarrow 0  ,\omega )]
 \nonumber \\
 &=& 0
\end{eqnarray}  
defines the kaon mass in the medium. 

Here we would like to state that, ideally, one should use the same model to 
compute the properties of the hadronic  matter as well as those of the kaon. 
In absence of a comprehensive model which would describe the nuclear matter 
properties as well as the kaon-baryon interaction, one is forced to employ two 
different models as we have done. However, we must stress that it would be 
incorrect to add the two Lagrangians ( ${\cal L}_{NK}$ and ${\cal L}_W$ ) and 
solve the problem since these two model Lagrangians are developed for two 
different purposes. ${\cal L}_{NK}$ has a $SU(3) \times SU(3)$ symmetry 
whereas ${\cal L}_W$ does not. Furthermore, at a conceptual level, such an 
approach  will amount to serious overcounting. For example, the $\sigma$ meson 
of the Walecka model Lagrangian essentially arises from the interacting 
two-pion system. On the other hand, the interactions of pseudoscalar mesons 
are explicitly included in the Nelson-Kaplan model.  
 
The behaviour of the kaon mass in the nuclear medium is displayed in Fig. 2. 
In this figure, the mean field results as well as the results of the full 
calculation are plotted. The results are for $a_3 m_s = -134 MeV$. We have also
shown the results when the hyperons are excluded. The calculation shows that
the loop diagrams tend to increase the kaon mass for $a_3 m_s = -134 MeV$. 
In fact, we find that the contribution of P-wave interaction ( Fig. 1(c) ) is 
negligible. One reason for this is that the contribution of this diagram is 
proportional to the square of the kaon momentum; in the static limit, 
therefore, this diagram contributes very little. We have used different 
parameter sets of the Walecka model in our calculations. In particular, the 
nuclear compressibility has been varied from 250 MeV to 350 MeV. We find that 
the kaon mass does not depend much on the compressibility. 

The effect of the inclusion of hyperons on the kaon mass is also displayed in
Fig. 2. We find that the reduction in the kaon mass is somewhat smaller when 
the hyperons are included. Thus, the density at which the kaon mass vanishes 
is increased when the hyperons are included and therefore the threshold for 
kaon condensation goes up with the inclusion of the hyperons. Of course, the 
hyperons ( primarily $\Sigma^-$'s  and $\Lambda$'s ) appear when the nuclear 
density is about 0.4 $fm^{-3}$. Hence the curves for with and without hyperons 
are identical below this density. Qualitatively, this behaviour of the kaon 
mass can be explained as follows. At higher densities, the energy would be 
minimised when the strangeness fraction is about unity. For example, the quark 
matter would have strangeness fraction of unity if the difference between 
strange and nonstrange quark masses is small in comparison with the quark 
chemical potential. In the hadronic matter, the strangeness fraction is 
increased by introducing hyperons or by having kaon condensate. Thus the 
presence of hyperons means that the system already has some strangeness 
fraction and therefore the kaon condensation is inhibited. We further find 
that the variation in the meson-baryon coupling strength does not appreciably 
affect the effective kaon mass. This result is somewhat at variance with the 
observation of earlier authors \cite{g}. The reason for this difference is 
most probably the naive addition of ${\cal L}_{NK}$ and ${\cal L_{W}}$ in ref.
\cite{g}.

The dependence of the kaon mass on $a_3m_s$,{\it i.e.}, the strangeness content 
of the nucleon, is shown in Fig. 3 at the mean field level. We find that the 
kaon mass decreases more rapidly when $a_3m_s$ decreases from -134 MeV to -310 
MeV. Thus, with the decrease in  $a_3m_s$ the threshold for kaon condensation 
decreases. A similar trend has been observed by other authors\cite{g}. 
However, we believe the smaller values of $a_3m_s$ are probably not 
meaningful, although there is some justification for these values in terms of 
strangeness content of the nucleon and pion-nucleon $\Sigma$ term\cite{k}. The 
reason is that at the lower values of $a_3m_s$, the kaon mass is as small as 
350 MeV when the nuclear density is 0.1 $fm^{-3}$, which is about two-thirds 
of nuclear density. Clearly, this would have a significant impact on the 
behaviour of kaons in nuclei. We are not aware of any experimental evidence 
which indicates such small $K^-$ masses.

Fig. 4 is plotted to study the effect of $a_{3}m_{s}$ on the higher order 
diagram. We find that higher order diagram disfavours the kaon condensation 
for higher values of $a_{3}m_{s}$ as effective kaon mass vanishes at higher 
densities. But for lower values of $a_{3}m_{s}$ oyster diagram favours the 
condensation. In other words, the higher order diagrams in the kaon-baryon 
interaction favours the condensation for higher strangeness fraction of the 
nucleon. 

In conclusion, we have shown that in the presence of hyperons kaon mass 
vanishes at higher density compared to the case where there is no hyperon in 
the medium. The effective kaon mass is more sensitive to the value of 
$a_3 m_s$ than the different parameter sets in the Walecka model which we have 
used here. The higher order diagrams in S-wave K-N interaction causes the kaon 
mass to vanish at lower densities for lower values of $a_3 m_s$. Here we would 
like to point out that a lower value of $a_3 m_s$ corresponds to higher 
strangeness fraction in nucleon. If the strangeness fraction is higher than 
$20\%$ \cite{k} then the kaon mass may be zero even at normal nuclear matter 
densities. This implies that kaon condensation may occur at normal nuclear 
matter or even at lower densities which, needless to say, is ruled out by 
physical arguments. One possible way out may be to incorporate the effects of 
finite size of mesons and baryons which may increase the critical density for 
baryons substantially. Studies in such direction are in progress. 

AB would like to thank Department of Atomic Energy (Government of India) and 
SKG would like to thank CSIR for support.
\pagebreak

%\begin{figure}[h]
%\epsfxsize=12cm
%\centerline{\epsfbox{fig1.eps}}
%\caption{The diagrams included in the calculation. }
%\end{figure}

%\pagebreak

%\begin{figure}[h]
%\epsfxsize=12cm
%\centerline{\epsfbox{fig2.ps}}
%\vspace{3in}
%\caption{Plot of kaon mass {\it vs.} baryon density. The graphs are for 
%$a_3m_s$ = -134 MeV. }
%\end{figure}

%\pagebreak

%\begin{figure}[h]
%\epsfxsize=12cm
%\centerline{\epsfbox{fig3.ps}}
%\vspace{2in}
%\caption{Plot of the kaon mass {\it vs.} baryon density. The graphs are for
%the calculation with all the graphs of Fig. 1 and with hyperons. }
%\end{figure}

\vspace{1.0cm}
%\end{document}

\end{document}